\documentclass[12pt,preprint]{aastex}
\usepackage{graphicx}
\usepackage{lscape}
\begin{document}
\title{Infrared Laboratory Oscillator Strengths of Fe I in the H-Band}
\author{M. P. Ruffoni, J. C. Pickering}
\affil{Blackett Laboratory, Dept. Physics, Imperial College London, London SW7 2AZ, UK}
\email{m.ruffoni@imperial.ac.uk}
\author{C. Allende Prieto}
\affil{Instituto de Astrof\'isica de Canarias, E38205 La Laguna, Tenerife, Spain}
\author{G. Nave}
\affil{National Institute of Standards and Technology, Gaithersburg, Maryland 20899-8422, USA}
\begin{abstract}
We report experimental oscillator strengths for 28 infrared Fe~I transitions, for which no previous experimental values exist. These transitions were selected to address an urgent need for oscillator strengths of lines in the H-band (between $1.4~\mu$m and $1.7~\mu$m) required for the analysis of spectra obtained from the Sloan Digital Sky Survey (SDSS-III) Apache Point Galactic Evolution Experiment (APOGEE). Upper limits have been placed on the oscillator strengths of an additional 7 transitions, predicted to be significant by published semi-empirical calculations, but not observed to be so. 
\end{abstract}


\section{Introduction}\label{section:intro}
In recent years, infrared (IR) spectroscopy has become an important tool for stellar astronomy, as demonstrated by results from the Spitzer and Wide-field Infrared Survey Explorer (WISE) space telescopes, the Cryogenic High-Resolution Infrared Echelle Spectrograph (CRIRES) at the Very Large Telescope, and the Radial Velocity Experiment (RAVE) spectrometer at the Siding Spring Observatory, to name but a few. This research has been made possible by the continual development of better IR detectors for both ground-based and satellite-borne spectrometers; a trend set to continue in the coming decade with the European Space Agency's (ESA) Gaia mission, and planned spectrometers for the European Extremely Large Telescope (E-ELT) facility.

However, during these studies it has become clear that the analysis of expensively acquired astrophysical spectra is often not limited by the capabilities of the spectrometers themselves, but by the lack of accurate reference data in the atomic line database (see \cite{ref:boeche08} and \cite{ref:bigot06}, for example). One project presently being affected in this way is the Apache Point Galactic Evolution Experiment (APOGEE) - part of the third Sloan Digital Sky Survey (SDSS-III) \citep{ref:eisenstein11,ref:cap08,2010IAUS..265..480M}.

APOGEE is using a highly-multiplexed near-IR spectrometer \citep{ref:eisenstein11,2012SPIE.8446E..0HW}, capable of acquiring spectra in the H-Band (between $1.4 \mu$m and $1.7 \mu$m) from up to 300 stars simultaneously with a spectral resolving power of the order of 30,000. At these wavelengths Galactic dust absorption is more than 5 times lower than in the optical (e.g. $A_H / A_V = 0.16$), allowing stars in the Galactic bulge and disc to be probed in addition to other stellar populations. During the planned three years of bright time a homogenous data set of 100,000 stars will be obtained, providing an integrated picture of the chemical and kinematical evolution of the Galaxy \citep{ref:cap08}.

Among the stellar properties being derived from these spectra are the abundances of 15 elements, including Fe peak elements, which will be measured to a precision of 0.1 dex (the unit dex stands for decimal exponent. $x$ dex $=10^x$). A key requirement for such measurements is the availability of accurate oscillator strengths, f (usually used as the $\log(g_lf)$, where $g_l$ is the statistical weight of the lower level), for strong enough, well-resolved lines seen within the H-Band. 

However, in the important case of neutral iron (Fe I), the recent critical compilation of $\log(g_lf)$s by \cite{ref:fuhr06} listed only the 51 transitions of \cite{ref:obrian} with wavelength greater than 1 $\mu$m; grading them C to D+, indicating uncertainties in $\log(g_lf)$ of 25\% to 50\%.

This inadequacy is at least in part due to the difficulty in measuring $\log(g_lf)$s in the infrared. Typically, oscillator strengths are obtained in the laboratory from measurements of atomic transition probabilities, A \citep{ref:spectrophysicsB}.
\begin{equation}
\log(g_lf) = \log\Bigl[A_{ul}g_u \lambda^2\times 1.499\times 10^{-14} \Bigr]~,
\end{equation}

where the subscript $u$ denotes a target upper energy level, and $ul$, a transition from this level to a lower state, $l$, that results in emission of photons of wavelength $\lambda$ (nm). $g_u$ is the statistical weight of the upper level. The $A_{ul}$ values are found by combining experimental branching fractions, $\mbox{BF}_{ul}$, with radiative lifetimes, $\tau_u$ \citep{ref:huber86}.
\begin{equation}
\label{eqn:trprob}
A_{ul} = \frac{\mbox{BF}_{ul}}{\tau_u}~;~~ \tau_u = \frac{1}{\sum_l A_{ul}}.
\end{equation}
The $\mbox{BF}_{ul}$ are commonly obtained from the ratio of calibrated emission line intensities, measured, for example, with Fourier transform spectrometery \citep{ref:pickering01}. This was the technique adopted in this study, and is described in more detail in Section \ref{section:bf}. However, for reliable line intensity calibration, all emission lines from an upper level must be sufficiently closely spaced in wavenumber to either be observed in a single spectrum, or be seen in a number of overlapping spectra where at least one line from the target upper level is present in the overlap region to carry the intensity calibration from one spectrum to the next.

$\tau_u$ is usually measured with time-resolved laser induced fluorescence (LIF). In a typical LIF experiment, such as that reported by \cite{ref:obrian} and \cite{ref:engelke}, an atomic beam of the target material is generated by a hollow-cathode discharge lamp (HCL), and is then excited by a light pulse from a pumped dye laser, the frequency of which is tunable in the visible to UV range. For a target upper level to be populated by this pulse, a strong transition must exist between it and an already populated lower level, where the energy difference matches that of the laser light. 

However, when studying infrared transitions, the requirements for each of these two experiments often become mutually exclusive. BFs can be obtained for transitions from upper levels that only produce closely-spaced infrared lines, but since these levels are relatively highly excited, they are not easily populated in a LIF experiment. Conversely, LIF measurements can be performed on upper levels that include at least one transition to a populated low-lying state, but since such a transition would usually be seen in the ultraviolet, a large gap would be present between it and any infrared lines, preventing emission spectra acquired in the two regions from being placed on a common intensity scale.

In identifying the Fe I transitions that will be required for the analysis of APOGEE spectra, we have found that the former situation typically prevails. We have therefore measured BFs in the normal way \citep{ref:pickering01}, as described in Section \ref{section:bf}, but have been forced to seek alternative means of normalising them to obtain the required transition probabilities from Equation \ref{eqn:trprob}. 

In Section \ref{section:lifetimes}, we describe a method for obtaining `effective lifetimes' for target upper levels by refining the semi-empirical lifetimes of \cite{ref:kurucz07}. Transition probabilities obtained with reference to these effective lifetimes are then examined further in section \ref{section:bowtie}, where we employ the long-established Ladenburg technique to obtain \emph{relative} $\log(g_lf)$s from networks of interconnected transitions without prior knowledge of the associated upper level lifetimes \citep{ref:ladenburg33}.

Results obtained from each of these two methods are then presented in Section \ref{section:results}, where we also provide recommended $\log(g_lf)$ values for the most critical Fe I transitions needed for the APOGEE project. Further transitions of interest to APOGEE (and complementary projects, such as Gaia-ESO) will be published in the near future.

\section{Experimental Procedure}\label{section:expt}

\subsection{Branching fraction measurements}
\label{section:bf}
The branching fraction of a given transition from upper level, $u$, is the ratio of its $A_{ul}$ to the sum of all $A_{ul}$ associated with $u$. Given an intensity calibrated line spectrum, this is equivalent to the ratio of observed relative line intensities for these transitions.

\begin{equation}
\label{eqn:bf}
\mbox{BF}_{ul} = \frac{A_{ul}}{\sum_lA_{ul}} = \frac{I_{ul}}{\sum_lI_{ul}}
\end{equation}


This approach does not depend on any form of equilibrium in the population distribution over different levels, but it is essential that all significant transitions from $u$ be included in the sum over $l$.

The BFs reported here were obtained from Fe~I atomic emission line spectra measured between $1800$ cm$^{-1}$ and $25000$ cm$^{-1}$ (between $5555$ nm and $400$ nm) on the 2 m Fourier transform (FT) spectrometer at the National Institute of Standards and Technology (NIST) \citep{ref:xgremlin}. The spectrum was excited from a 99.8\% pure Fe cathode mounted in a water cooled HCL running under the conditions shown in Table \ref{table:spectra}. All the results reported in Section \ref{section:results} were obtained from spectrum A, with spectra B and C being used to check the target lines were free from self-absorption or blends with carrier gas lines.

The intensity of each spectrum was calibrated using a tungsten (W) halogen lamp with spectral radiance known to $\pm$1.1\% between $250$ nm and $6000$ nm. Spectra of this lamp were recorded both before and after the acquisition of each Fe spectrum using the same spectrometer parameters, and the measured intensity used to obtain the spectrometer response function. The full procedure is discussed in detail by \cite{ref:pickering01}. The number of Fe scans taken between each W lamp spectrum was limited to ensure all three spectra (W, Fe, W) could be acquired within approximately 2 hours. Over this period, negligible variation in spectrometer response was observed. The two W lamp spectra were thus co-added, and the resulting spectrum used to obtain the spectrometer response function, from which the Fe spectrum was intensity calibrated.

Where additional Fe scans were required to increase the signal-to-noise (S/N) ratio of target transitions, multiple sets of W, Fe, W spectra were obtained, and each of the calibrated Fe spectra co-added to produce a single composite spectrum.

\begin{deluxetable}{lcccl}
\tablewidth{0pt}
\tabletypesize{\scriptsize}
\tablecaption{FT spectra recorded for the BF measurements}
\tablehead{
\colhead{Tag} & \colhead{Wavenumber}        & \colhead{Carrier Gas and} & \colhead{Current} & Spectrum Filename(s)\tablenotemark{a} \\
\colhead{}         & \colhead{Range (cm$^{-1}$)} & \colhead{Pressure (mbar)} & \colhead{(A)}
}
\startdata
A & 1800-15000 & Ne, 3.7 & 2.0 & Fe072511.002 to .010; Fe072611.006 to .012; Fe072711.002 to .005\\
B & 1800-15000 & Ne, 3.7 & 1.4 & Fe072511.001 \\
C & 1800-15000 & Ar, 2.7 & 2.0 & Fe072201.001 to .004\\
\enddata
\tablecomments{All spectra were measured with InSb detectors on the 2 m FT spectrometer at NIST.}
\tablenotetext{a}{Where more than one filename is given, the named spectra were coadded to improve the signal-to-noise ratio of weak lines.}
\label{table:spectra}
\end{deluxetable}

\subsection{Line Fitting}
\begin{figure}
\includegraphics[angle=270,scale=0.65]{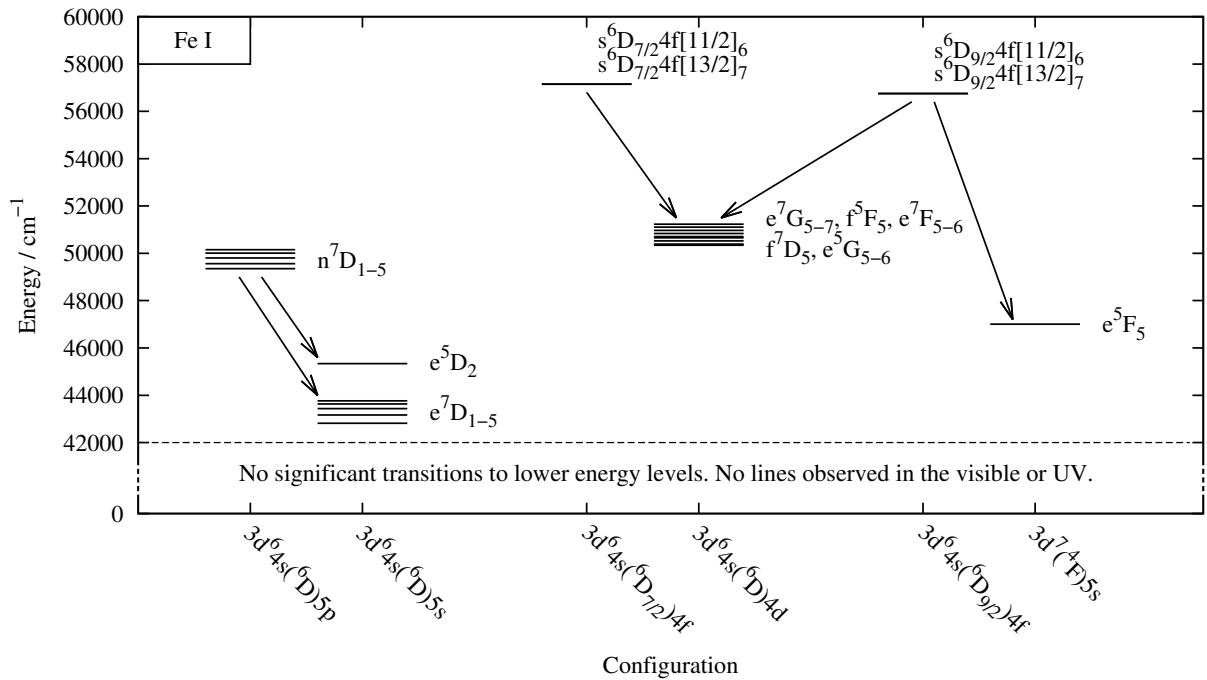}
\caption{The Fe I upper energy levels for which branching fractions to the indicated lower levels were measured. Each level, or collection of levels where they are close together in energy, is labelled with its term and grouped with others of the same configuration.} 
\label{fig:energy}
\end{figure}

For each target upper level, the semi-empirical calculations of \cite{ref:kurucz07} were used to give a list of predicted transitions to lower levels, as shown diagrammatically in Figure \ref{fig:energy}. The emission lines of these transitions were then identified in our Fe spectra, and the \verb|XGremlin| package \citep{ref:xgremlin} used to obtain relative line intensities by fitting Voigt profiles to them. Some lines, predicted by Kurucz to contribute less than 1 \% to the total BF, were too weak to be observed. The sum total of their predicted branching fractions was thus assigned to an unobserved `residual'. This was then included in the summation over $l$ in equation \ref{eqn:bf} to normalise the sum of all BFs to unity. BFs and $\log(g_lf)$s were obtained with the aid of the \verb|FAST| package \citep{ref:ruffoni12}.

In some cases, there was disagreement between the predicted intensity of a line and that observed experimentally. Where lines were predicted to be visible above our experimental noise limit, but were absent in our spectra, we assigned a maximum possible BF to each unobserved line based on the noise at its expected wavenumber. These disagreements were associated with transitions from $4f$ upper levels, and would likely be due to a high degree of level mixing, lowering the accuracy of predicted line intensities.

\subsection{Uncertainties in Branching Fractions and $\log(g_lf)$s}
\label{section:uncert}
Several sources of uncertainty contributed to the overall uncertainty in intensity of a given line:
\begin{enumerate}
\item{The uncertainty in intensity of the line profile, obtained from the inverse of its signal-to-noise ratio (S/N).}
\item{The uncertainty in intensity of the tungsten standard lamp when measured on the FT spectrometer, obtained from the inverse of its S/N at the same wavenumber as the line.}
\item{The uncertainty in the calibrated spectral radiance of the tungsten lamp at the same wavenumber as the line. Within one spectrum we ascribe $1/\sqrt{2}$ of the quoted one standard deviation lamp calibration error, $\epsilon$, to each line so that the uncertainty in the intensity ratio of any two lines is $\epsilon$.}
\end{enumerate}

These are added in quadrature to give an overall uncertainty in the calibrated intensity of a given line profile, $\Delta I_{ul}$. Uncertainties in individual BFs, $\Delta(BF)_{ul}$, are then given by
\begin{equation}
\Biggl(\frac{\Delta (BF)_{ul}}{(BF)_{ul}}\Biggr)^2 = (1 - 2(BF)_{ul})\Biggl(\frac{\Delta I_{ul}}{I_{ul}}\Biggr)^2 + \sum_{j=1}^n(BF)_{uj}^2\Biggl(\frac{\Delta I_{uj}}{I_{uj}}\Biggr)^2~.
\end{equation}

This formalism is derived from Equation 7 in \cite{ref:sikstrom02}, and takes special account of the correlation between $I_{ul}$ and $\sum_l I_{ul}$ that arises from the presence of a given line intensity in both the numerator and denominator on the right-hand side of Equation \ref{eqn:bf}. The uncertainty in the transition probability of a given line, $\Delta A_{ul}$, is then
\begin{equation}
\Biggl(\frac{\Delta A_{ul}}{A_{ul}}\Biggr)^2 = \Biggl(\frac{\Delta (BF)_{ul}}{(BF)_{ul}}\Biggr)^2 + \Biggl(\frac{\Delta \tau_{ul}}{\tau_{ul}}\Biggr)^2~,
\end{equation}

where $\Delta \tau_{ul}$ is in the uncertainty in upper level lifetime. This then leads to the uncertainty in $\log(g_lf)$ of a given line,
\begin{equation}
\Delta \log(g_lf) (\mbox{dex}) = \log\Biggl[g_lf \Biggl(1 +  \frac{\Delta A_{ul}}{A_{ul}}\Biggr)\Biggr] - \log[g_lf]~,
\end{equation}

\subsection{Determining Effective Radiative Lifetimes}
\label{section:lifetimes}
Upper level radiative lifetimes are typically obtained from LIF measurements, as described in Section \ref{section:intro}. However, this technique cannot be applied to the levels studied here, forcing us to adopt a different approach.

Table \ref{table:lifetimes} lists the upper levels from which we wish to find the $\log(g_lf)$s of transitions to lower levels. The Theory column shows the upper level lifetimes calculated by \cite{ref:kurucz07}. These were combined with our experimental BFs to arrive at an initial estimate to the $\log(g_lf)$ of each transition listed in Table \ref{table:linefit}. These $\log(g_lf)$s were then used in conjunction with a 1D model atmosphere interpolated from the calculations by \cite{ref:castelli04} for the atmospheric parameters of the Sun ($T_{eff} = 5777$ K, log(gravity in cm s$^{-2}$) = 4.437) to generate  synthetic Fe I spectral lines, which were then compared with the solar spectrum from the atlas by \cite{ref:livingston91} (see also \cite{ref:wallace96}). This solar atlas was obtained with the Fourier Transform Spectrograph formerly at the Pierce-McMath facility at Kitt Peak National Observatory, with a resolving power $R \sim 300,000$.

The spectral synthesis calculations were performed with the ASS$\epsilon$T code \citep{ref:koesterke08, ref:koesterke09}. We would have liked to have obtained an optimized solar iron abundance from fits to Fe I H-band lines with known, accurate $\log(gf)$s. However, as already stated in the motivation for this study, such data are not available. In their absence, we adopted the solar abundance of $7.45 \pm 0.05$ dex\footnote{On the usual logarithmic abundance scale, where the hydrogen abundance is 12.00 by definition.} from \cite{ref:asplund05}. Although this value was obtained by those authors through the use of a 3D stellar atmosphere code, 3D effects on Fe I lines in the Sun are known to be small \citep{ref:asplund00, ref:cap02}. Furthermore, other recent solar studies using 1D model atmospheres, have produced Fe abundances in agreement with this value \citep{ref:ramirez13}.
 
We made use of the most complete line list available to us, including both atomic and molecular transitions (M. Shetrone, private communication), which facilitates the analysis of Fe I lines in the vicinity of other transitions. Radiative and Stark damping constants were adopted from Kuruczs website while Van der Waals (VdW) while damping constants (collisional damping due to neutral atoms) were taken from \cite{ref:melendez99} when
available (calculated from the tables by \cite{ref:anstee95}, \cite{ref:barklem97} and \cite{ref:barklem98}) and slightly adjusted to match the solar spectrum \citep{ref:shetrone}.

A micro-turbulence of 1.1 km s$^{-1}$ was included in the radiative transfer calculations. Each model spectrum was convolved with a Gaussian with a full-width at half maximum (FWHM) of 2.3 km s$^{-1}$ to account for the instrumental profile (FWHM $\sim$ 1 km s$^{-1}$) and, most importantly, macro-turbulence. The adopted macro-turbulence is smaller than the values typically found in the optical (see, e.g. \cite{ref:cap01}). Such difference is likely related to the Fe I infrared lines forming in a different range of atmospheric layers than Fe I optical lines due to the lower continuum opacity in the H band. Wavelength offsets were removed (since it  was line intensities that were of interest), and the background continuum level adjusted when  necessary. The intensity of each model line, I$_{model}$, was then compared to the corresponding line intensity observed, I$_{obs}$, and our initial $\log(g_lf)$ varied until the two were matched. This provided a $\log(g_lf)$ correction factor, $\delta\log(g_lf)$, for each line.

\begin{figure*}
\centering
\includegraphics[trim=0 0 0 13cm, clip=true, totalheight=0.5\textheight, scale=0.4]{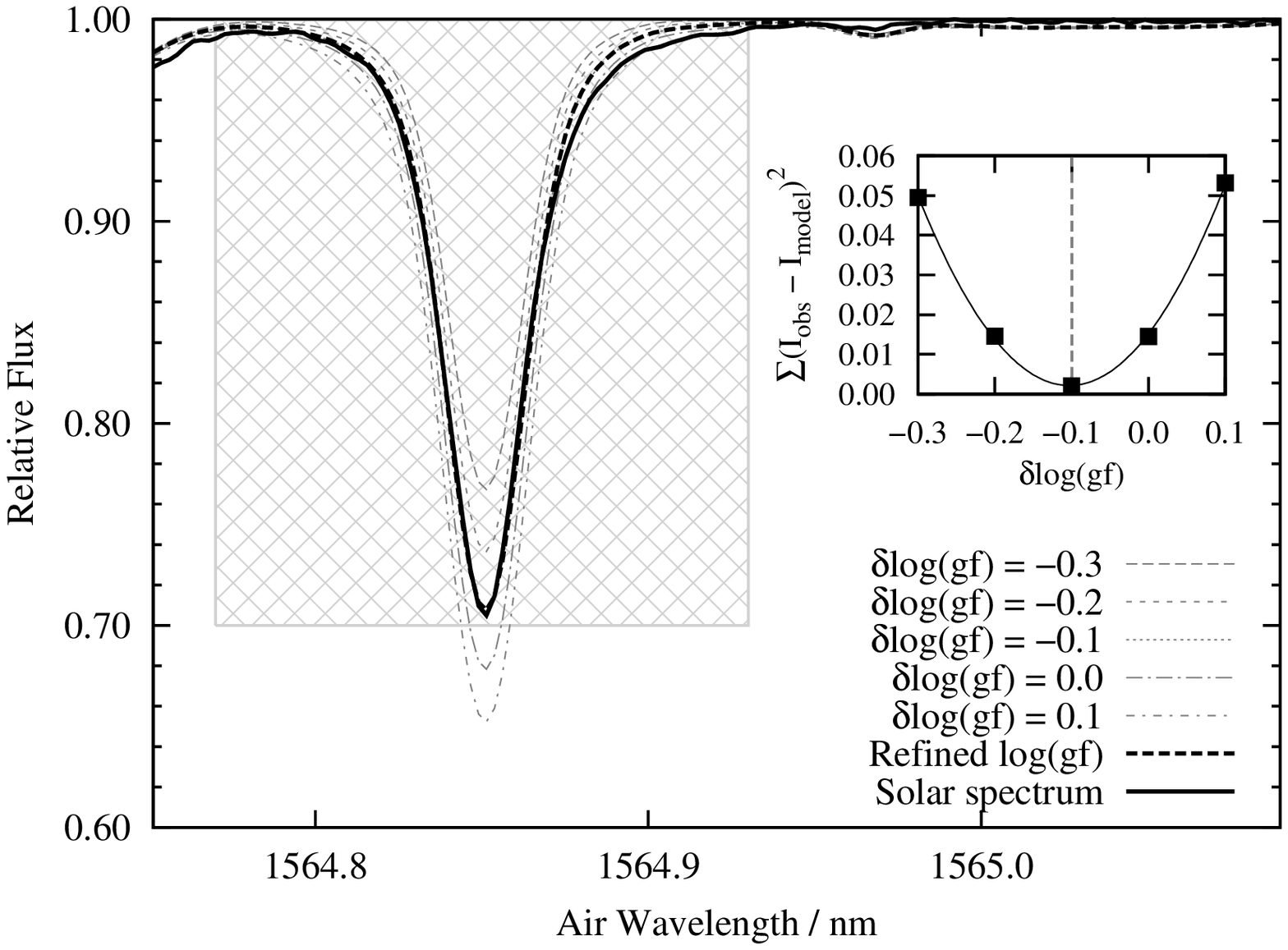}
\includegraphics[trim=0 0 0 13cm clip=true, totalheight=0.5\textheight, scale=0.4]{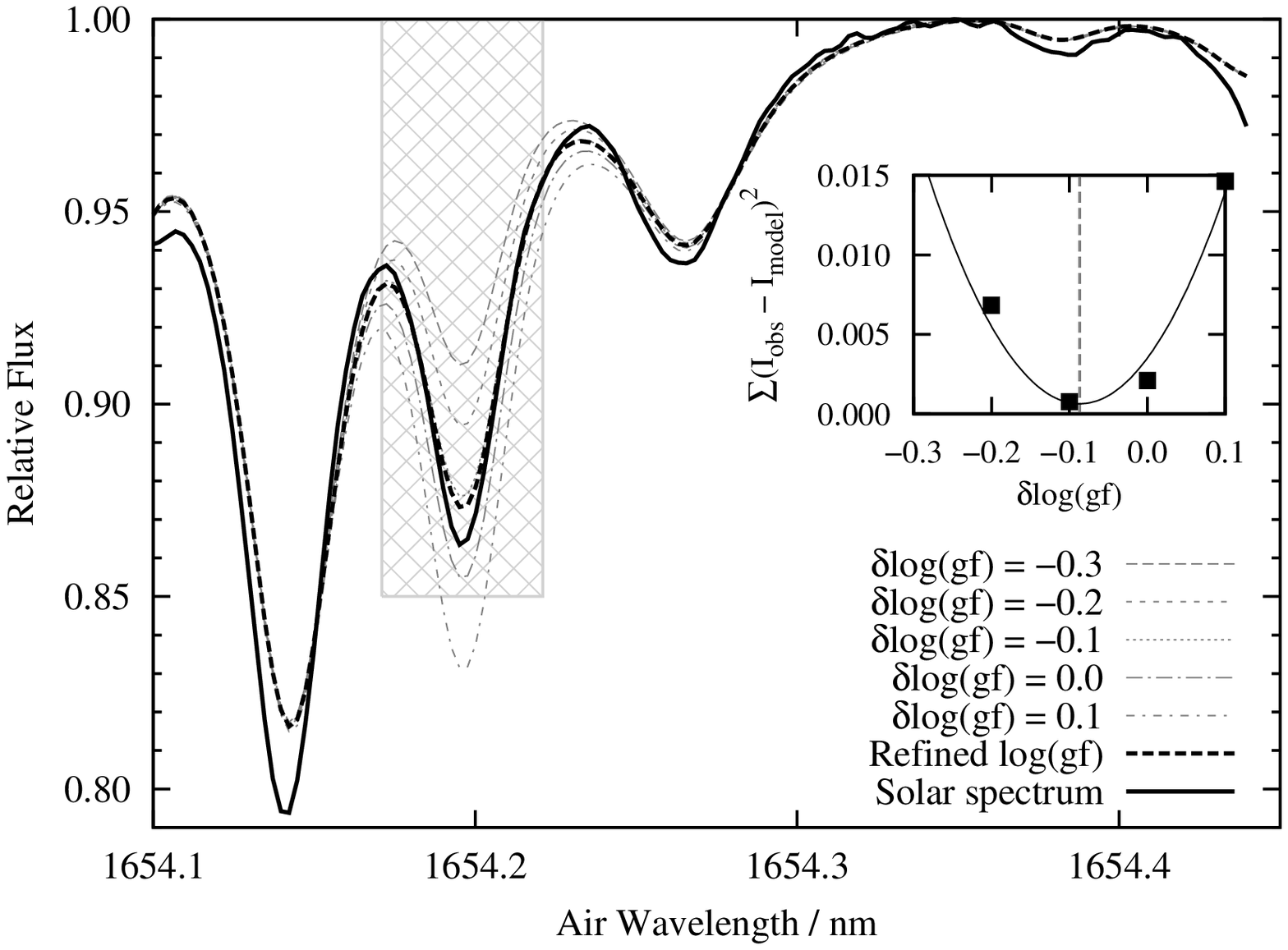}
\caption{Determining $\delta\log(g_lf)$s from fits of modelled line profiles to stellar spectra.}
\label{fig:linefit}
\end{figure*}

Figure \ref{fig:linefit} shows two examples of the fitted line profiles. Our refined synthetic profile is plotted as a dashed black line and the observed solar profile as a solid black line. The remaining solid and dashed grey plots show the change in I$_{model}$ for successive $\pm 0.1$ offsets to our initial $\log(g_lf)$.  The cross-hatched grey area around each line shows the region used to compare the model and observed spectra. Its width was chosen to avoid neighbouring lines, and its height restricted to relative fluxes between 0.7 and 1.0 to limit the influence of possible non-local thermodynamic equilibrium (NLTE) effects.  NLTE effects tend to be most important in the outermost, low-density atmospheric layers, and therefore affect most significantly the cores of strong lines. The inset in each plot shows the sum of differences in relative flux, $(I_{obs} - I_{model})^2$, as a function of $\delta\log(g_lf)$, where the dashed vertical line indicates the optimal value of $\delta\log(g_lf)$, obtained from the minimum of a parabolic fit to the differences at each $\delta\log(g_lf)$. Table \ref{table:linefit} lists all the lines for which refined $\log(g_lf)$s were determined.

If it is assumed that each $\delta\log(g_lf)$ is due to an error in the corresponding Kurucz upper level lifetime, rather than any other source, it can be directly related to a correction in $\tau_u$ . By adding each $\delta$ $\log(g_lf)$ to its initial experimental $\log(g_lf)$ and determining the corresponding value of A$_{ul}$, Equation \ref{eqn:trprob} can be inverted to obtain an improved estimate of the upper level lifetime, as listed in Table \ref{table:linefit}. Averaging these values over all lines belonging to a given upper level then provides a refined value of $\tau_u$ for that level, as listed in Table \ref{table:lifetimes} in the ``Effective Lifetime" column. These were then combined with our experimentally measured BFs to provide refined $\log(g_lf)$s for all lines from each target upper level, as described in Section \ref{section:results} and listed in ``BF \& Effective $\tau$" column in Table \ref{table:results}.

The uncertainties in the individual lifetimes shown in Table \ref{table:linefit} were estimated by combining the uncertainty in BF (and thus intensity) for the line, the uncertainty in determining $\delta\log(g_lf)$, and the standard deviation of refined lifetimes for all lines belonging to the same upper level. When these were combined to provide the uncertainties in effective lifetimes listed in Table \ref{table:lifetimes}, an additional uncertainty of 12\% was included to account for the $0.05$ dex uncertainty in solar Fe abundance quoted by \cite{ref:asplund05}.

This uncertainty in solar Fe abundance significantly increased the overall uncertainty in the effective lifetimes. Yet nonetheless, the refinement procedure reduced the probable error in each lifetime from approximately 20\% or more, found by taking the median difference between the \cite{ref:kurucz07} lifetimes and experimental values for the levels studied in \cite{ref:obrian} and \cite{ref:engelke}, to between 12\% and 15\%.

\begin{deluxetable}{lccccc}
\tablewidth{0pt}
\tabletypesize{\scriptsize}
\tablecaption{Radiative lifetimes for Fe I levels linked to transitions of high importance to APOGEE.}
\tablehead{
\colhead{} & \colhead{} & \colhead{} & \colhead{} & \multicolumn{2}{c}{Lifetime, $\tau$ (ns)} \\
\colhead{Configuration} & \colhead{Term} & \colhead{J} & \colhead{Energy (cm$^{-1}$)} & Theory\tablenotemark{a} & Effective\tablenotemark{b}
}
\startdata
$3d^64s(^6D)5p$       & $n^7D$ & $5$ & $49352.338$ & $71.429$ & $97 \pm 12$ \\
                      & $n^7D$ & $4$ & $49558.731$ & $69.930$ & $101 \pm 14$ \\
                      & $n^7D$ & $3$ & $49805.254$ & $68.027$ & $95 \pm 12$ \\
                      & $n^7D$ & $2$ & $50008.519$ & $68.493$ & $94 \pm 14$ \\
                      & $n^7D$ & $1$ & $50152.616$ & $71.429$ & $89 \pm 11$ \\
$3d^64s(^6D_{7/2})4f$ & $s^6D_{3.5}4f[5.5]$ & $6$ & $57146.768$ & $55.249$ & $67 \pm 9$ \\
                      & $s^6D_{3.5}4f[6.5]$ & $7$ & $57152.331$ & $56.818$ & $69 \pm 18$ \\
$3d^64s(^6D_{9/2})4f$ & $s^6D_{4.5}4f[5.5]$ & $6$ & $56748.897$ & $46.512$ & $41 \pm 6$ \\
                      & $s^6D_{4.5}4f[6.5]$ & $7$ & $56754.128$ & $54.945$ & $54 \pm 7$ \\
\enddata
\tablecomments{The configuration, term, and energy level data are taken from \cite{ref:nave94}.}
\tablenotetext{a}{Semi-empirical values calculated by \cite{ref:kurucz07}.}
\tablenotetext{b}{The uncertainty in effective lifetimes includes the 0.05 dex uncertainty in the solar Fe abundance of \cite{ref:asplund05}}
\label{table:lifetimes}
\end{deluxetable}

\begin{deluxetable}{lcrrcccc}
\tablewidth{0pt}
\tabletypesize{\scriptsize}
\tablecaption{Effective upper level lifetimes found for individual transitions in stellar spectra}
\tablehead{
\colhead{Upper}       & \colhead{Lower}       & \multicolumn{2}{c}{Transition\tablenotemark{a}} & \colhead{Initial}                   & \colhead{Stellar}      & \colhead{VdW Broadening} & \colhead{Effective} \\
\colhead{Level ($u$)} & \colhead{Level ($l$)} & \colhead{Air $\lambda_{ul}$ (nm)}               & \colhead{$\sigma_{ul}$ (cm$^{-1}$)} & \colhead{$\log(g_lf)$} & \colhead{$\delta\log(g_lf)$} & \colhead{(rad s$^{-1}$cm$^3$)} & \colhead{$\tau_u$ (ns)}

}
\startdata
$n^7D_5$ & $e^7D_4$ & $1615.3249$ & $6189.014$ & $-0.529$ & $-0.149$ & $-6.97$ & $101 \pm 4$\\
         & $e^7D_5$ & $1529.4562$ & $6536.485$ & $0.710$  & $-0.117$ & $-7.21$ & $94 \pm 4$ \\
$n^7D_4$ & $e^7D_3$ & $1632.4459$ & $6124.104$ & $-0.414$ & $-0.136$ & $-6.96$ & $94 \pm 8$ \\
         & $e^7D_4$ & $1563.1950$ & $6395.407$ & $0.186$  & $-0.201$ & $-6.96$ & $109 \pm 8$ \\
$n^7D_3$ & $e^7D_2$ & $1619.8505$ & $6171.723$ & $-0.317$ & $-0.156$ & $-6.95$ & $97 \pm 3$ \\
         & $e^7D_3$ & $1569.2751$ & $6370.628$ & $-0.369$ & $-0.133$ & $-6.95$ & $92 \pm 3$ \\
$n^7D_2$ & $e^7D_1$ & $1600.9615$ & $6244.540$ & $-0.375$ & $-0.114$ & $-6.95$ & $91 \pm 14$ \\
         & $e^7D_2$ & $1568.2021$ & $6374.987$ & $-2.200$ & $-0.036$ & $-7.15$ & $76 \pm 20$ \\
         & $e^7D_3$ & $1520.7530$ & $6573.893$ & $0.307$  & $-0.192$ & $-7.02$ & $109 \pm 14$ \\
$n^7D_1$ & $e^7D_1$ & $1564.8515$ & $6388.637$ & $-0.516$ & $-0.103$ & $-6.75$ & $91 \pm 2$ \\
         & $e^7D_2$ & $1533.5387$ & $6519.084$ & $0.070$  & $-0.090$ & $-7.07$ & $88 \pm 1$ \\
$s^6D_{3.5}4f[5.5]_6$ 
         & $f^5F_5$ & $1654.1968$ & $6043.579$ & $-0.253$ & $-0.087$ & $-6.75$ & $68 \pm 5$ \\
         & $e^7G_6$ & $1617.9585$ & $6178.940$ & $0.227$  & $-0.106$ & $-6.75$ & $71 \pm 3$ \\
         & $e^7F_5$ & $1583.5164$ & $6313.334$ & $0.790$  & $-0.054$ & $-7.01$ & $63 \pm 3$ \\
$s^6D_{3.5}4f[6.5]_7$
         & $e^7G_6$ & $1616.5031$ & $6184.503$ & $0.969$  & $-0.170$ & $-6.78$ & $84 \pm 16$ \\
         & $e^7F_6$ & $1467.9844$ & $6810.200$ & $-0.137$ &  $0.032$ & $-6.80$ & $53 \pm 16$ \\
$s^6D_{4.5}4f[5.5]_6$ 
         & $e^5G_5$ & $1653.8000$ & $6045.029$ & $-0.639$ &  $0.096$ & $-6.89$ & $37 \pm 7$ \\
         & $e^7G_7$ & $1639.6311$ & $6097.267$ & $-0.683$ &  $0.099$ & $-6.80$ & $37 \pm 7$ \\
         & $f^7D_5$ & $1569.1857$ & $6370.991$ & $0.543$  &  $0.005$ & $-6.80$ & $46 \pm 4$ \\
$s^6D_{4.5}4f[6.5]_7$ 
         & $e^7G_7$ & $1638.2256$ & $6102.498$ & $0.342$  &  $0.000$ & $-6.80$ & $55 \pm 2$ \\
         & $e^7F_6$ & $1559.1497$ & $6412.000$ & $0.887$  &  $0.020$ & $-7.00$ & $53 \pm 1$ \\
\enddata 
\tablenotetext{a}{Transition wavenumber and air wavelength data from \cite{ref:nave94}.}           
\label{table:linefit}
\end{deluxetable}

\subsection{Verifying the Refined $\log(g_lf)$s using the Ladenburg Technique}
\label{section:bowtie}

Given the non-standard approach we have been forced to adopt to obtain effective upper level lifetimes, the accuracy of the refined $\log(g_lf)$s described in Section \ref{section:lifetimes} must be examined further.

Relative $\log(g_lf)$s can be obtained for a set of transitions originating from a common upper or lower level by comparing their relative calibrated line intensities \citep{ref:huber86,ref:spectrophysics,ref:ladenburg33}. When these line intensities are measured in emission, upper level branching \emph{ratios} are obtained, which are similar to the branching \emph{fractions} described in Section \ref{section:bf}, but do not necessarily include all significant transitions from $u$ to $l$. When measured in absorption, relative $\log(g_lf)$s can be found for transitions to different upper levels that share a common lower level. Thus, a network of linked upper and lower levels may be formed. The relative $\log(g_lf)$s of all the interconnecting transitions are placed on an absolute scale by finding the absolute $\log(g_lf)$ of any one transition. 

Whilst the accuracy of these $\log(g_lf)$s relies on the accuracy of the chosen reference $\log(g_lf)$, this approach has the advantage that not all transitions down from an upper level, or up from a lower level, must be included in the network. Furthermore, and most importantly from the perspective of verifying the accuracy of the refined $\log(g_lf)$s reported in Table \ref{table:results}, no knowledge of the upper level lifetime is required.

\begin{figure*}
\centering
\includegraphics[scale=0.32]{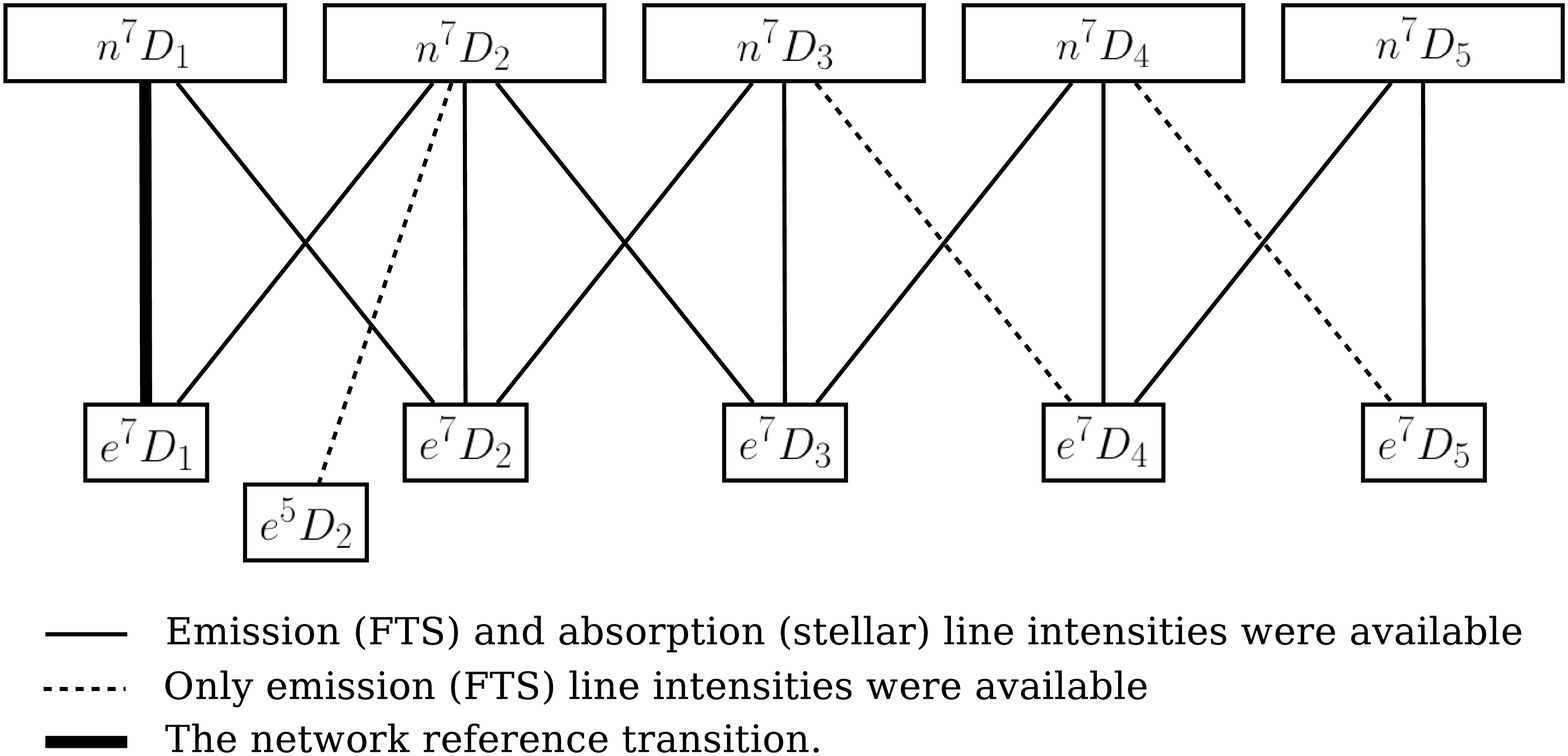}
\caption{The network of transitions that link the studied $n^7D$ upper levels.}
\label{fig:bowtie_n7d}
\end{figure*}

Figure \ref{fig:bowtie_n7d} shows the transitions that link the five $n^7D$ upper levels studied here. The effective lifetimes for the two transitions from $n^7D_1$ show the best consistency in Table \ref{table:linefit}, with the transition from $n^7D_1$ to $e^7D_1$ also exhibited the smallest $\delta\log(g_lf)$ after refining the Kurucz lifetime. We therefore asserted that its refined experimental $\log(g_lf)$ is a good representation of the absolute $\log(g_lf)$ that would have been obtained if $\tau_u$ were known. This was thus the reference transition against which we measured relative $\log(g_lf)$s of all the other transitions in the network.

Table \ref{table:bowtie_n7D} lists the calibrated line intensities for this network of transitions, measured in both emission, with FT spectroscopy, and absorption, from our synthetic line profile fits to the solar spectrum of \cite{ref:livingston91}. The $\log(g_lf)$ of the $n^7D_1$ to $e^7D_1$ transition was fixed to the refined value listed in Table \ref{table:results}. The Transfer Ratio column then shows which two levels were used to link this $\log(g_lf)$ to others in the network. Where this ratio is between two lower levels in the network, the transfer was made using the ratio of emission line intensities from a common upper level. Where it is between two upper levels, the transfer was made using the ratio of absorption line intensities from a common lower level.

The Ladenburg $\log(g_lf)$ of each transition and its uncertainty are listed in Table \ref{table:results}, where it can be seen that there is close agreement with the refined $\log(g_lf)$s obtained by combining branching fractions with effective upper level lifetimes. However, in some cases the Ladenburg $\log(g_lf)$s have a lower uncertainty than those found by combining BFs with an effective $\tau_u$. In these instances, we recommend the use of the Ladenburg $\log(g_lf)$.

\begin{figure*}
\centering
\includegraphics[scale=0.32]{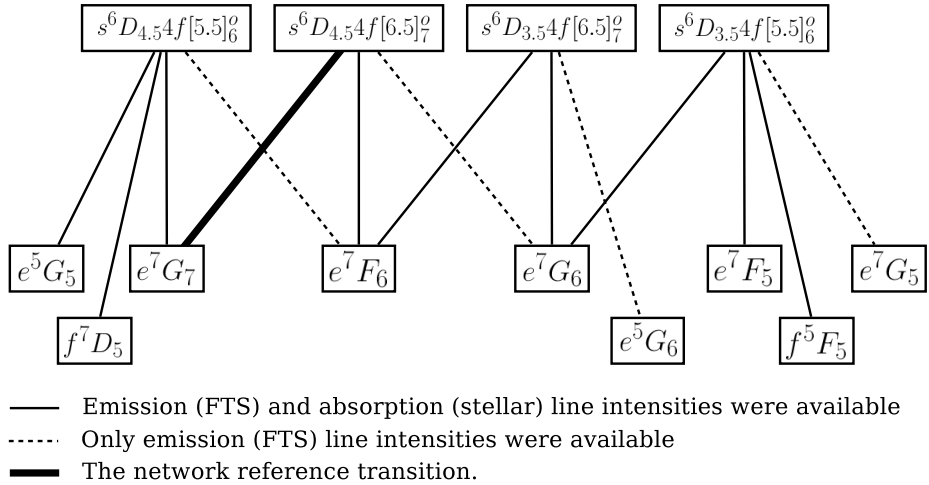}
\caption{The network of transitions that link the studied $4f$ upper levels.}
\label{fig:bowtie_4f}
\end{figure*}

 
Figure \ref{fig:bowtie_4f} and Table \ref{table:bowtie_4f} show similar results for the $4f$ upper levels studied here. In this network, the $s^6D_{4.5}4f[6.5]_7$ to $e^7G_7$ transition was chosen to be the reference transition as its $\log(g_lf)$ was unchanged by the refinement process in Section \ref{section:lifetimes}. The Ladenburg $\log(g_lf)$s listed in Table \ref{table:results} again agree with the refined $\log(g_lf)$s


\newpage

\begin{landscape}
\begin{deluxetable}{lcrrccc}
\tablewidth{0pt}
\tabletypesize{\scriptsize}
\tablecaption{The network of linked transitions stemming from the $n^7D_1$ to $e^7D_1$ transition.}
\tablehead{
\colhead{Upper}     & \colhead{Lower}     & \multicolumn{2}{c}{Transition}                                           & \colhead{Emission}   & \colhead{Absorption} & \colhead{Transfer} \\
\colhead{Level (u)} & \colhead{Level (l)} & \colhead{Air $\lambda_{ul}$ (nm)} & \colhead{$\sigma_{ul}$ (cm$^{-1}$)} & \colhead{$I$ (arb. unit)} & \colhead{$I$ (arb. unit)} & \colhead{Ratio}

}
\startdata
$n^7D_5$ & $e^7D_4$ & $1615.3249$ & $6189.014$ & $8.765 \times 10^3$ & $4.879 \times 10^5$ & $n^7D_5 / n^7D_4$ \\
           & $e^7D_5$ & $1529.4562$ & $6536.485$ & $1.696 \times 10^5$ & $$                  & $e^7D_5   / e^7D_4$   \\
$n^7D_4$ & $e^7D_3$ & $1632.4459$ & $6124.104$ & $1.036 \times 10^4$ & $7.679 \times 10^5$ & $n^7D_4 / n^7D_3$ \\
           & $e^7D_4$ & $1563.1950$ & $6395.407$ & $4.499 \times 10^4$ & $2.930 \times 10^6$ & $e^7D_4   / e^7D_3$   \\
           & $e^7D_5$ & $1482.6412$ & $6742.877$ & $8.356 \times 10^4$ & $$                  & $e^7D_5   / e^7D_3$   \\
$n^7D_3$ & $e^7D_2$ & $1619.8505$ & $6171.723$ & $1.229 \times 10^4$ & $1.223 \times 10^6$ & $n^7D_3 / n^7D_1$ \\
           & $e^7D_3$ & $1569.2751$ & $6370.628$ & $1.162 \times 10^4$ & $1.218 \times 10^6$ & $e^7D_3   / e^7D_2$   \\
           & $e^7D_4$ & $1505.1749$ & $6641.931$ & $7.555 \times 10^4$ & $$                  & $e^7D_4   / e^7D_2$   \\
$n^7D_2$ & $e^5D_2$ & $2138.6166$ & $4674.644$ & $2.589 \times 10^2$ & $$                  & $e^5D_2   / e^7D_1$   \\
           & $e^7D_1$ & $1600.9615$ & $6244.540$ & $1.060 \times 10^4$ & $1.688 \times 10^6$ & $n^7D_2 / n^7D_1$ \\
           & $e^7D_2$ & $1568.2021$ & $6374.987$ & $1.653 \times 10^2$ & $$                  & $e^7D_2   / e^7D_1$   \\
           & $e^7D_3$ & $1520.7530$ & $6573.893$ & $5.652 \times 10^4$ & $$                  & $e^7D_3   / e^7D_1$   \\
$n^7D_1$ & $e^7D_1$ & $1564.8515$ & $6388.637$ & $7.520 \times 10^3$ & $2.184 \times 10^6$ & Fixed                 \\
           & $e^7D_2$ & $1533.5387$ & $6519.084$ & $3.020 \times 10^4$ & $9.030 \times 10^6$ & $e^7D_2 / e^7D_1$     \\
\enddata            
\label{table:bowtie_n7D}
\end{deluxetable}

\begin{deluxetable}{lcrrccc}
\tablewidth{0pt}
\tabletypesize{\scriptsize}
\tablecaption{The network of linked transitions stemming from the $s^6D_{4.5}4f[6.5]_7$ to $e^7G_7$ transition.}
\tablehead{
\colhead{Upper}     & \colhead{Lower}     & \multicolumn{2}{c}{Transition}                                           & \colhead{Emission}   & \colhead{Absorption} & \colhead{Transfer} \\
\colhead{Level (u)} & \colhead{Level (l)} & \colhead{Air $\lambda_{ul}$ (nm)} & \colhead{$\sigma_{ul}$ (cm$^{-1}$)} & \colhead{$I$ (arb. unit)} & \colhead{$I$ (arb. unit)} & \colhead{Ratio}

}
\startdata
$s^6D_{3.5}4f[6.5]_7$
           & $e^7G_6$ & $1616.5031$ & $6184.503$ & $1.948 \times 10^4$ & $1.232 \times 10^7$ & $e^7G_6 / e^7F_6$ \\
           & $e^5G_6$ & $1508.0225$ & $6629.389$ & $2.695 \times 10^2$ & $$                  & $e^5G_6 / e^7F_6$ \\
           & $e^7F_6$ & $1467.9844$ & $6810.200$ & $1.847 \times 10^3$ & $1.285 \times 10^6$ & $s^6D_{3.5}4f[6.5]_7 / s^6D_{4.5}4f[6.5]_7$ \\
$s^6D_{4.5}4f[6.5]_7$
           & $e^7G_6$ & $1727.7482$ & $5786.300$ & $5.194 \times 10^2$ & $$                  & $e^7G_6 / e^7G_7$ \\
           & $e^7G_7$ & $1638.2256$ & $6102.498$ & $4.718 \times 10^3$ & $3.642 \times 10^6$ & Fixed \\
           & $e^7F_6$ & $1559.1497$ & $6412.000$ & $1.829 \times 10^4$ & $1.477 \times 10^7$ & $e^7F_6 / e^7G_7$             \\
$s^6D_{4.5}4f[5.5]_6$ 
           & $e^5G_5$ & $1653.8000$  & $6045.029$ & $5.575 \times 10^2$ & $$                  & $e^5G_5 / e^7G_7$ \\
           & $e^7G_7$ & $1639.6311$ & $6097.267$ & $5.129 \times 10^2$ & $4.978 \times 10^5$ & $s^6D_{4.5}4f[5.5]_6 / s^6D_{4.5}4f[6.5]_7$ \\
           & $f^7D_5$ & $1569.1857$ & $6370.991$ & $9.423 \times 10^3$ & $$                  & $f^7D_5 / e^7G_7$ \\
           & $e^7F_6$ & $1560.4225$ & $6406.770$ & $1.263 \times 10^4$ & $$                  & $e^7F_6 / e^7G_7$ \\
$s^6D_{3.5}4f[5.5]_6$ 
           & $e^7G_5$ & $1689.2375$ & $5918.214$ & $5.409 \times 10^2$ & $$                  & $e^7G_5 / e^7G_6$ \\
           & $f^5F_5$ & $1654.1968$ & $6043.579$ & $1.024 \times 10^3$ & $$                  & $f^5F_5 / e^7G_6$ \\
           & $e^7G_6$ & $1617.9585$ & $6178.940$ & $3.230 \times 10^3$ & $2.591 \times 10^6$ & $s^6D_{3.5}4f[5.5]_6 / s^6D_{3.5}4f[6.5]_7$ \\
           & $e^7F_5$ & $1583.5164$ & $6313.334$ & $1.232 \times 10^4$ & $$                  & $e^7F_5 / e^7G_6$ \\
\enddata            
\label{table:bowtie_4f}
\end{deluxetable}
\end{landscape}

\newpage
\section{Results}
\label{section:results}

Measured branching fractions and oscillator strengths for transitions from the upper levels listed in Table \ref{table:lifetimes} are shown in Table \ref{table:results}. To our knowledge, no previous experimental $\log(g_lf)$s exist for transitions from these levels. The listed level identifications, wavenumbers, and air wavelengths were taken from \cite{ref:nave94}, where possible. In a small number of cases, \cite{ref:nave94} do not report a transition predicted to be significant by \cite{ref:kurucz07}. In these cases, the transition wavenumbers and wavelengths are shown in italics and were taken from \cite{ref:kurucz07}.

For easy comparison, Table \ref{table:results} lists the $\log(g_lf)$s obtained both by combining branching fractions with effective upper level lifetimes, and by employing the Ladenburg technique. The values of $A_{ul}$ shown correspond to the ``BF \& Effective $\tau$" $\log(g_lf)$ values. The final two columns list the $\log(g_lf)$ values that we recommend for use in future spectral analyses. They are those obtained by combining branching fractions with effective upper level lifetimes, except in cases where the Ladenburg technique provided  $\log(g_lf)$s with a lower experimental uncertainty.


\begin{landscape}
\begin{deluxetable}{lcrrrrrrrrrrr}
\tablewidth{0pt}
\tabletypesize{\scriptsize}
\tablecaption{Experimental branching fractions, transition probabilities, and $\log(g_lf)$s for the Fe~I levels listed in Table \ref{table:lifetimes}.}
\tablehead{
\colhead{Upper} & \colhead{Lower} & \multicolumn{2}{c}{Transition\tablenotemark{a}} & \colhead{BF} & \colhead{$\Delta$BF} & \colhead{$A_{ul}$} & \multicolumn{2}{c}{BF \& Effective $\tau$} & \multicolumn{2}{c}{Ladenburg} & \multicolumn{2}{c}{Recommended} \\
\colhead{Level (u)} & \colhead{Level (l)} & \colhead{Air $\lambda_{ul}$ (nm)} & \colhead{$\sigma_{ul}$ (cm$^{-1}$)} & & \colhead{(\%)}& \colhead{($10^6$s$^{-1}$)} & \colhead {$\log(g_lf)$} & \colhead{$\pm$ dex} & \colhead{$\log(g_lf)$} & \colhead{$\pm$ dex} & \colhead{$\log(g_lf)$} & \colhead{$\pm$ dex}

}
\startdata
$n^7D_5$   & $e^7D_4$ & $1615.3249$ & $6189.014$ & $0.049$ & $1.29$ & $0.506$ & $-0.66$ & $0.05$ & $-0.65$ & $0.06$ & $-0.66$ & $0.05$ \\
           & $e^7D_5$ & $1529.4562$ & $6536.485$ & $0.950$ & $0.07$ & $9.795$ &  $0.58$ & $0.05$ & $ 0.59$ & $0.06$ &  $0.58$ & $0.05$ \\
           &          &             & Residual   & $0.001$ \\
$n^7D_4$   & $e^7D_3$ & $1632.4459$ & $6124.104$ & $0.073$ & $1.12$ & $0.727$ & $-0.58$ & $0.06$ & $-0.59$ & $0.05$ & $-0.59$ & $0.05$ \\
           & $e^7D_4$ & $1563.1950$ & $6395.407$ & $0.319$ & $0.72$ & $3.156$ &  $0.02$ & $0.06$ & $ 0.01$ & $0.05$ & $ 0.01$ & $0.05$ \\
           & $e^7D_5$ & $1482.6412$ & $6742.877$ & $0.592$ & $0.42$ & $5.862$ &  $0.24$ & $0.06$ & $ 0.24$ & $0.05$ & $ 0.24$ & $0.05$ \\
           &          &             & Residual   & $0.016$ \\
$n^7D_3$   & $e^7D_2$ & $1619.8505$ & $6171.723$ & $0.119$ & $1.02$ & $1.254$ & $-0.46$ & $0.05$ & $-0.48$ & $0.05$ & $-0.46$ & $0.05$ \\
           & $e^7D_3$ & $1569.2751$ & $6370.628$ & $0.113$ & $1.18$ & $1.185$ & $-0.51$ & $0.05$ & $-0.53$ & $0.05$ & $-0.51$ & $0.05$ \\
           & $e^7D_4$ & $1505.1749$ & $6641.931$ & $0.732$ & $0.27$ & $7.708$ &  $0.26$ & $0.05$ & $ 0.25$ & $0.05$ & $ 0.26$ & $0.05$ \\
           &          &             & Residual   & $0.036$ \\
$n^7D_2$   & $e^5D_2$ & $2138.6166$ & $4674.644$ & $0.004$ & $24.6$& $0.040$  & $-1.86$ & $0.11$ & $-1.84$ & $0.11$ & $-1.84$ & $0.11$ \\
           & $e^7D_1$ & $1600.9615$ & $6244.540$ & $0.154$ & $1.04$ & $1.634$ & $-0.50$ & $0.06$ & $-0.48$ & $0.05$ & $-0.48$ & $0.05$ \\
           & $e^7D_2$ & $1568.2021$ & $6374.987$ & $0.002$ & $18.6$& $0.025$  & $-2.33$ & $0.09$ & $-2.31$ & $0.09$ & $-2.31$ & $0.09$ \\
           & $e^7D_3$ & $1520.7530$ & $6573.893$ & $0.819$ & $0.23$ & $8.711$ &  $0.18$ & $0.06$ & $ 0.20$ & $0.05$ & $ 0.20$ & $0.05$ \\
           &          &             & Residual   & $0.021$ \\
$n^7D_1$   & $e^7D_1$ & $1564.8515$ & $6388.637$ & $0.198$ & $1.18$ & $2.222$ & $-0.61$ & $0.05$ & $-0.61$ & $0.05$ & $-0.61$ & $0.05$ \\
           & $e^7D_2$ & $1533.5387$ & $6519.084$ & $0.794$ & $0.30$ & $8.925$ & $-0.03$ & $0.05$ & $-0.03$ & $0.05$ & $-0.03$ & $0.05$ \\
           &          &             & Residual   & $0.008$ \\
$s^6D_{3.5}4f[5.5]_6$ & $e^7G_5$ & $1689.2375$ & $5918.214$ & $0.031$ & $15.0$ & $0.457$  & $-0.60$ & $0.08$ & $-0.59$ & $0.08$ & $-0.60$ & $0.08$ \\
                      & $f^5F_5$ & $1654.1968$ & $6043.579$ & $0.058$ & $6.20$ & $0.865$  & $-0.34$ & $0.06$ & $-0.33$ & $0.06$ & $-0.34$ & $0.06$ \\
                      & $e^7G_6$ & $1617.9585$ & $6178.940$ & $0.183$ & $1.68$ & $2.727$  &  $0.14$ & $0.06$ & $ 0.15$ & $0.06$ &  $0.14$ & $0.06$ \\
                      & $e^7F_5$ & $1583.5164$ & $6313.334$ & $0.697$ & $0.81$ & $10.396$ &  $0.71$ & $0.05$ & $ 0.71$ & $0.06$ &  $0.71$ & $0.05$ \\
                      & $f^7D_5$ & $1476.9493$ & $6768.863$ & $<0.008$\tablenotemark{c} & $-$ & $0.120$ & $-1.29$ & $0.16$ & & & $-1.29$ & $0.16$ \\
                      &          &             & Residual   & $0.023$ \\
$s^6D_{3.5}4f[6.5]_7$ & $e^7G_6$ & $1616.5031$ & $6184.503$ & $0.901$ & $0.64$ & $12.692$ &  $0.87$ & $0.10$ & $ 0.89$ & $0.05$ & $ 0.89$ & $0.05$ \\
                      & $e^5G_6$ & $1508.0225$ & $6629.389$ & $0.012$ & $32.6$ & $0.176$  & $-1.05$ & $0.15$ & $-1.03$ & $0.13$ & $-1.03$ & $0.13$ \\
                      & $e^7F_6$ & $1467.9844$ & $6810.200$ & $0.085$ & $5.27$ & $1.204$  & $-0.23$ & $0.10$ & $-0.22$ & $0.05$ & $-0.22$ & $0.05$ \\
                      &          &             & Residual   & $0.002$ \\
$s^6D_{4.5}4f[5.5]_6$ & $e^7G_5$ & \emph{1810.986} & \emph{5520.347}\tablenotemark{b} & $<0.001$\tablenotemark{c} & $-$ & $0.013$ & $-2.09$ & $0.82$ & & & $-2.09$ & $0.82$ \\
                      & $f^7F_5$ & $1770.7734$ & $5645.708$ & $<0.005$\tablenotemark{c} & $-$ & $0.131$ & $-1.10$ & $0.11$ & & & $-1.10$ & $0.11$ \\
                      & $e^7F_5$ & $1690.0234$ & $5915.462$ & $<0.004$\tablenotemark{c} & $-$ & $0.091$ & $-1.30$ & $0.26$ & & & $-1.30$ & $0.26$ \\
                      & $e^5G_5$ & $1653.8000$ & $6045.029$ & $0.020$ & $14.0$ & $0.489$ & $-0.58$ & $0.08$ & $-0.53$ & $0.08$ & $-0.53$ & $0.08$ \\
                      & $e^7G_7$ & $1639.6311$ & $6097.267$ & $0.018$ & $16.4$ & $0.450$ & $-0.63$ & $0.09$ & $-0.58$ & $0.05$ & $-0.58$ & $0.05$ \\
                      & $e^5G_6$ & \emph{1605.740} & \emph{6225.956}\tablenotemark{b} & $<0.003$\tablenotemark{c} & $-$ & $0.063$ & $-1.50$ & $0.28$ & &  & $-1.50$ & $0.28$\\
                      & $f^7D_5$ & $1569.1857$ & $6370.991$ & $0.339$ & $1.80$ & $8.261$ & $0.60$ & $0.06$ & $ 0.65$ & $0.06$ & $0.65$ & $0.06$ \\
                      & $e^7F_6$ & $1560.4225$ & $6406.770$ & $0.454$ & $1.77$ & $11.068$ & $0.72$ & $0.06$ & $ 0.77$ & $0.06$ & $0.77$ & $0.06$ \\
                      & $e^5F_5$ & \emph{1026.055} & \emph{9743.394}\tablenotemark{b} & $<0.001$\tablenotemark{c} & $-$ & $0.013$ & $-2.56$ & $1.28$ & & & $-2.56$ & $1.28$ \\
                      & $a^5F_5$ &  \emph{200.655} & \emph{49820.629} & \emph{0.152}\tablenotemark{d} & $-$ \\
                      &          &             & Residual   & $0.003$ \\
$s^6D_{4.5}4f[6.5]_7$ & $e^7G_6$ & $1727.7482$ & $5786.300$ & $0.022$ & $20.0$ & $0.408$  & $-0.56$ & $0.09$ & $-0.56$ & $0.09$ & $-0.56$ & $0.09$ \\
                      & $e^7G_7$ & $1638.2256$ & $6102.498$ & $0.200$ & $1.92$ & $3.704$  &  $0.35$ & $0.05$ & $ 0.35$ & $0.05$ &  $0.35$ & $0.05$ \\
                      & $e^5G_6$ & \emph{1604.392} & \emph{6231.187}\tablenotemark{b} & $<0.002$\tablenotemark{c} & $-$ & $0.037$ & $-1.67$ & $0.18$ & & & $-1.67$ & $0.18$ \\
                      & $e^7F_6$ & $1559.1497$ & $6412.000$ & $0.776$ & $0.67$ & $14.362$ &  $0.90$ & $0.05$ & $ 0.90$ & $0.05$ &  $0.90$ & $0.05$ \\
                      &          &             & Residual   & $<0.001$ \\
\enddata
\tablecomments{The Residual BF is the sum total BF of all other lines predicted by Kurucz for the given upper level, but which were either outside the measurement range or predicted to contribute less than 1\% of the total BF.}
\tablenotetext{a}{Transition wavenumber and air wavelength data from \cite{ref:nave94}.}
\tablenotetext{b}{\cite{ref:nave94} did not detect these lines. Transition data shown is therefore from \cite{ref:kurucz07}.}
\tablenotetext{c}{Kurucz predicts these transitions have BFs of at least 1\%, yet they are unobserved in our spectra. For each transition, the quoted wavelength and wavenumber (in italics) are therefore those given by \cite{ref:kurucz07}, and the BF shown is an upper limit given the spectral noise in that region.}
\tablenotetext{d}{{L}ine outside the measured spectral range. BF predicted from data in \cite{ref:kurucz07}.}
\label{table:results}
\end{deluxetable}
\end{landscape}

\section{Summary}
Oscillator strengths have been found for 28 levels of Fe I in the H-band for which no experimental laboratory data have previously existed. Upper limits on possible $\log(g_lf)$ values have been assigned to an additional 7 transitions. Together these almost double the number of laboratory measured IR Fe I $\log(g_lf)$s available to astronomers in the atomic database. The uncertainty in each $\log(g_lf)$ value is between $0.05$ dex and $0.06$ dex for the stronger lines, and $0.08$ dex and $0.11$ dex for the majority of observed weak lines. This is sufficient to allow analysis of APOGEE spectra to reach the desired $0.1$ dex accuracy in determining chemical abundances.

The $\log(g_lf)$s presented here were derived from laboratory measured BFs combined with effective upper level lifetimes obtained from stellar spectra. Clearly, improvements could be made to these values in future with laboratory measured lifetimes, but this will require further development of existing LIF experiments, or the development of new lifetime measurement methods. One technique that holds promise in this respect would be the coupling of an IR laser frequency comb with time domain FT spectroscopy measurements.

However, in spite of this limitation, we have employed the Ladenburg technique of determining relative $\log(g_lf)$s for networks of related transitions to demonstrate that the $\log(g_lf)$s reported here are consistent relative to one another, and -- assuming the reported $\log(g_lf)$s for the $n^7D_1$ to $e^7D_1$ and $s^6D_{4.5}4f[6.5]_7$ to $e^7G_7$ transitions are correct -- are accurate on an absolute scale. If necessary, the uncertainties in the Ladenburg $\log(g_lf)$s may be reduced in the future by replacing the solar line intensities shown in Tables \ref{table:bowtie_n7D} and \ref{table:bowtie_4f} with laboratory-measured absorption intensities \citep{ref:cardon79}. This would eliminate, for example, non-LTE, convective, and magnto hydrodynamic effects that arise from the complex nature of the solar photosphere.

The results presented here represent the first part of an effort to provide the community with laboratory data for transition metal elements in the infra-red; particularly in the regions of importance to surveys such as APOGEE and Gaia-ESO, space-borne missions such as ESA's upcoming Gaia satellite, and to support future research at facilities such as E-ELT. Additional results will therefore be published in the near future.

\acknowledgments
We would like to thank T. Ryabchikova, E. A. Den Hartog, A. P. Thorne, and S. L. Redman for many helpful discussions; and M. Shetrone and D. Bizyaev for providing us with their VdW damping constants relating to the APOGEE linelist. MPR and JCP also thank the UK Science and Technology Facilities Council (STFC) for funding this research.

\clearpage

\end{document}